\documentstyle[useAMS,graphicx,longtable]{mn2e}

\newcommand{\hMpc}{{\ifmmode{h^{-1}{\rm Mpc}}\else{$h^{-1}$Mpc}\fi}}
\newcommand{\hkpc}{{\ifmmode{h^{-1}{\rm kpc}}\else{$h^{-1}$kpc}\fi}}

\def\approxlt{\mathrel{\spose{\lower 3pt\hbox{$\sim$}}
        \raise 2.0pt\hbox{$<$}}}
\def\approxgt{\mathrel{\spose{\lower 3pt\hbox{$\sim$}}
        \raise 2.0pt\hbox{$>$}}}
\def\approxpropto{\mathrel{\spose{\lower 3pt\hbox{$\sim$}}
        \raise 2.0pt\hbox{$\propto$}}}

\title{Gamma-Ray Bursts from quark stars}
\author[B. Paczy{\'n}ski and P. Haensel]
{B. Paczy{\'n}ski$^{1}$\thanks{e-mail: bp@astro.princeton.edu}
and P. Haensel$^{1}$\thanks{e-mail: haensel@camk.edu.pl}\\
$^{1}$Princeton University Observatory, Peyton Hall,
Princeton, NJ 08544\\
$^{2}$N. Copernicus Astronomical Center, Polish Academy of
Sciences, Bartycka 18, PL-00-716 Warszawa, Poland}

\date{Accepted 2005 May 5.
      Received 2005 May 4;
      in original form  2005 February 16}

\pubyear{2005}

\begin{document}

\maketitle

\label{firstpage}

\begin{abstract}

Long gamma-ray bursts (GRBs) are believed to be related to the
explosion of type Ic supernovae, which have been stripped of
their hydrogen and helium envelopes. There appear to be two
types of these explosions: those which are approximately
spherical (GRB980425/1998bw), and which are associated with
weak bursts, and the classical GRBs which generate
ultra-relativistic jets (GRB030329/SN2003dh). If this
bimodality is real {\it Swift} will provide a clear evidence
for it.

We propose that classical powerful GRBs, which generate
ultrarelativistic outflows, are a result of a formation of
quark stars.  Quark stars may provide an additional energy for
the explosion of SN Ic, but far more important is a creation
of a surface which acts as a membrane which cannot be
penetrated by baryons.  A surface of a quark star allows only
ultrarelativistic matter to escape: photons, neutrinos,
electron -- positron pairs and magnetic fields.  The formation
of a quark star follows several minutes after the initial core
collapse. Possible evidence for this time delay is provided by
BATSE precursors to GRBs, as analyzed by Lazzati (2005).

\end{abstract}

\begin{keywords}
Gamma-Ray Bursts -- stars: neutron -- supernovae -- stars: quark
\end{keywords}

\section{Introduction}
\label{sect:intro}

It is generally accepted that long gamma-ray bursts (GRBs) are
associated with star forming regions (Paczy\'nski 1998).  More
specifically, they are related to supernovae (SNe)of type Ic,
the core collapse events of massive stars stripped of their
hydrogen and helium envelopes (Filippenko 1997, Iwamoto et al.
1998, Heger et al. 2003, Stanek 2004, Stanek et al. 2005, and
references therein). While the SN/GRB connection was gradually
emerging, it was most clearly demonstrated by a recent
GRB030329, associated with SN2003dh (Stanek et al. 2003,
Vanderspek et al. 2004).

We propose that there are two types of SNe Ic associated with
GRBs.  One is a more or less spherical explosion as originally
described by Colgate (1968), prior to the discovery of the
first GRBs (Klebesadel at al. 1973).  The spherical model was
refined (Chevalier and Li 1999, Tan et al. 2001), and it may
be relevant to events like SN1998bw/GRB980425 (Galama et al.
1998, Iwamoto et al. 1998), and SN2001em (Stockdale et al.
2004, 2005; Bietenholtz et al. 2005). The other kind, with a
strong ultra-relativistic jet, is a classical long GRB, much
more powerful in gamma-rays than the first type. The proposed
bimodality, if real, will be verified with {\it Swift}.

We speculate about the physical reason for the difference
between the two types of supernovae.  We propose that the
strong ultrarelativistic outflows are developed when the
collapsing core becomes a quark star.  These hypothetical
stars  are built of self-bound quark matter (Witten 1984,
Haensel et al. 1986, Alcock et al. 1986). The total energy
released could range from $10^{52}$~erg (Cheng and Dai 1996)
to $10^{53}$~erg (Bombaci and Datta 2000), but we do not think
that neutron star -- strange star conversion is energetically
decisive for SN Ic.  However, the formation of a baryon --
quark membrane at the surface of the collapsed compact object
is most essential. The membrane can be penetrated by photons,
electrons, positrons, neutrinos and the magnetic fields, i.e.
by various forms of ultrarelativistic flow. { The idea of a
baryon -- quark membrane is  crucial for the model proposed in
this Letter. Such a membrane prevents baryon contamination of
the energy outflow from a hot quark star, a property already
used in a  model of short GRBs by Haensel et. (1991). }
Modeling of classical strong GRBs by energy outflow from newly
born quark star was previously suggested by several authors
(Dai and Lu 1998, Wang et al. 2000, Ouyed and Sannino 2002,
Berezhiani et al. 2003, Drago et al. 2004).

We shall outline the evolution to a supernova type Ic in the
next section, and we shall discuss the theoretical possibility
of the formation of a quark star in Section 3. We shall
conclude this letter with a discussion emphasizing the
decisive role that {\it Swift} (Gehrels et al. 2004) will play
in verifying our speculations.

\section{Evolution to a supernova} 
\label{sect:birth.SN}

The evolution leading to a formation of SN Ic is agreed upon
(cf. Lee et al. 2000, Della Valle 2005, and references
therein).  We need a massive star to be stripped of hydrogen
and helium envelopes. This may happen to a single Wolf-Rayet
(WR)  star, or to a WR star in a close binary system. In the
latter case the binary may be compact and tidally locked, with
a rapid rotation as a consequence.  When the star runs out of
nuclear fuel the inner core collapses, ejecting a fair amount
of ${\rm Ni}^{56}$, which is the dominant energy source for
SNe Ic (Iwamoto et al. 1998, Deng et al. 2005), with up to
${\rm 10^{52.5} }$ erg of kinetic energy.  The relativistic
outflow from a gamma-ray burst is estimated to be ${\rm
10^{51.5} }$ erg (Frail et al. 2000, Berger et al. 2004).
Therefore, the energetics of a GRB is only a relatively modest
fraction of the SN Ic explosion energy.  What makes it
extraordinary is the bulk Lorentz factor of the
ultrarelativistic component, estimated to be ${\rm \Gamma \sim
100 - 1000 }$ (Meszaros et al. 1998).

As far as we can tell,  the evolution as described in the last
paragraph is not controversial.  What is controversial is the
nature of a collapsed object: is it a neutron star or a black
hole, and even more controversially: what are physical
processes capable of ejecting bulk matter at ${\rm \Gamma \sim
300}$? The common suggestions are neutrino -- antineutrino
annihilation (Woosley 1993), and magnetic fields (Usov 1992,
Klu\'zniak and Ruderman 1998, Lee et al. 2000, Blandford 2002,
Lyutikov and Blandford 2004, and references therein). The
details are very complicated, and it is not possible to prove
or disprove those models.  The large bulk Lorenz factor is
obtained by postulating suitable initial conditions, i.e. by
replacing one puzzle with another.

Note that black holes are not a necessary condition for the
formation of jets.  Models of magnetized neutron stars by
LeBlanc and Wilson (1970) demonstrated that they can form
jets; even more importantly it has been found observationally
that Crab and other compact stars do form jets (Mori et al.
2004, and references therein). Also note that neutron stars
are formed out of stars with the initial mass as high as 50
${\rm M}_\odot$ (Kaper et al. 1995, Clark et al. 2002, Figer
et al. 2005, Gaensler et al. 2005).

If a quark star is formed its surface separates the outside,
where baryons can exist, from the inside where baryons are dissolved
into quarks.  The surface of a quark star separates
baryonic matter from non-baryonic, it creates a membrane which may
may be crossed only by electrons and positrons, photons,
neutrino pairs, and magnetic fields, automatically generating
conditions needed for an ultra-relativistic outflow of a
classical GRB.  There is a price to be paid:
we do not know if quark stars exist.

At this stage a theory is too complicated or too uncertain,
but we may ask: how far can we proceed using semi-empirical
evidence?

Classical, powerful GRBs are believed to be jet-like, and if we are
not in their beam we shall not see a burst.
Perhaps the anomalously weak GRB980425 was
due to a jet which missed us.  However, no evidence for a jet
several years after the explosion makes this
scenario unlikely (Waxman 2004),
and there is an emerging consensus that this was not a classical GRB
(Soderberg et al. 2004).

Following the suggestion by Paczy\'nski (2001) and Granot and
Loeb (2003) radio observations were done with Very Large Array
(VLA) (Stockdale et al. 2004, Bietenholtz and Bartel 2005) and
Very Long Baseline Array (VLBA) (Stockdale et al. 2005) to
monitor an SN of type Ic, SN2001em. While strong radio
emission was clearly detected, there was no evidence of a
relativistic jet. It appears to be the second case of a SN Ic
explosion which is not associated with a misaligned
relativistic GRB jet.

Some SNe Ic are clearly associated with relativistic GRB jets,
like GRB030329/SN2003dh (Stanek et al. 2003, Vanderspek et al.
2004)), for which the superluminal jet moving at (3-5)$c$ was
observed by Taylor et al. (2004). Another good case of a
GRB/SN association was provided by GRB031203/SN2003lw
(Malesani et al. 2004).  In addition several GRB afterglows
have very pronounced "bumps" in their light curves, providing
photometric evidence of the underlying supernovae (Stanek et
al. 2005, Zeh 2005).

While this is only a preliminary conclusion it appears that
the powerful SNe Ic come in two types: those which generate
classical GRBs with ultra-relativistic jets, and those which
do not generate ultra-relativistic GRB jets.  Note, the GRB
events without relativistic jets are natural (Colgate 1968).
Its principle is very simple: a spherical explosion generates
a shock wave which accelerates outer stellar shells to
sub-relativistic velocity and generates a weak gamma-ray
burst.  What is extraordinary is the formation of powerful
ultra-relativistic jets in some of these SN Ic.  What is the
reason for this diversity, while the corresponding supernovae
appear to be similar? We suggest that the difference is due to
formation of a core made of self-bound quark matter (SQM) in
some of these explosive events.  But we should keep in mind,
however, the possibility that the bimodality of SNe Ic is due
to the presence or absence of rotation.

Our suggestion of a bimodal distribution of SN Ic properties
is based on a small statistics: two and two events of each
type. This may be improved with {\it Swift}. According to
Norris (2002, Fig. 5) there is a population of about 90 BATSE
bursts which are relatively nearby, with a typical distance of
$ \sim $ 100 Mpc, which show a concentration to super galactic
plane. They seem to be related to GRB980425.  If some of these
GRBs were jet-like, but we were not in the beam, we should be
able to detect their radio emission several years later
(Paczy\'nski 2001, Granot and Loeb 2003).  The proposed
emission from a decelerated jet was not detected in 1998bw
(Soderberg et al. 2004) or in 2001em (Stockdale 2004, 2005;
Bietenholtz and Bartel 2005), but with a much larger number of
BATSE GRBs the radio emission should be detectable in at least
some cases. The large BATSE error circles will not be a
problem according to van Gorkom (2005) and Frail (2005), as
the radio sky shows relatively little variability, and the
late GRBs should stand out above the background.

Note that long-lag GRBs contribute only $ \sim 5 $ percent  at
the bright end of BATSE luminosity function, but they dominate
with 50 percent at the faint end (cf. Norris 2002, Fig 3, 4).
Their number counts are "Euclidean", i.e. they are definitely
nearby, at least by cosmological standards. {\it Swift}
(Gehrels et al. 2004) will provide GRB coordinates accurate to
an arcminute and even arcsecond within a minute of the event.
If the distribution discovered by Norris (2002) holds then
{\it Swift} will provide GRB/SN relation for many events and
it will strengthen (or weaken) our proposal that some SN Ic
explosions are more or less spherical (Colgate 1968), and some
generate ultra-relativistic jets (the classical GRBs).

\section{Formation of a quark star}
\label{sect:birth.QS}

As we have already mentioned in Section 1, the total energy
$10^{52} - 10^{53}$~erg released in the neutron star -- quark
star conversion is not very important for the energetics of
SNe Ic.  { In our scenario the conversion energy is emitted
after the supernova shock breakout,   mostly in neutrinos.}
However, if quark stars are formed then they are essential as
a source of an ultra-relativistic outflow. A quark star
surface acts as a membrane confining quarks to the stellar
volume, which constitutes a huge bubble (bag) of the quantum
chromodynamics  vacuum. The membrane allows for absorption of
baryons by SQM, where they dissolve into quarks. At
temperatures under consideration, this membrane may be crossed
only by electrons and positrons, photons, neutrino pairs and
magnetic fields, automatically generating conditions needed
for an ultrarelativistic outflow of classical GRBs.

The concept of an outflow from a compact star collimated into
a narrow jet has been around for decades (LeBlanc and Wilson
1970, Mori et al. 2004).  It is not to claim that the problem
has been theoretically solved, but the relativistic outflow
from a hypothetical quark star does not create any new
problems.  In fact,  the relativistic outflow should be made
easier with the generation of baryon free ejection.

We do not propose to solve the collimation problem, as this is
common to all GRB models, and the presence of jets is
supported by numerous observations (cf. Stanek et al. 1999,
and reference therein). We do not propose a novel solution for
a SN explosion; we know that SNe explode. The issue we propose
to solve with quark stars is the generation of
ultrarelativistic outflows.

We consider the formation of a hot quark star at the SN Ic
centre. The pre-SN Fe-Ni core collapses into a proto-neutron
star. Structure and evolution of proto-neutron stars has been
recently studied by Pons et al. (1999, 2001a, 2001b), and its
dynamics by Villain et al. (2004) and Ferrari et al. (2003).
Neutrinos trapped in dense matter enforce its high electron
fraction ($\sim 30$ percent), stiffen the equation of state,
and prohibit the formation of hyperons. All these effects
result in a relatively low central density ($\sim 5\times
10^{14} ~{\rm g~cm^{-3}}$), and a large radius of a
proto-neutron star ($\sim 50$ km). However, within a minute or
so an excess of electron neutrinos diffuses out, and the
electron fraction in the core falls to the usual $\sim 10$
percent. This deleptonization allows for the appearance of
hyperons, which additionally softens the equation of state.
The central density rises to $\sim 10^{15}~{\rm g~cm^{-3}}$,
and temperature rises to some $5\times 10^{11}$~K: a hot
neutron star is born.

Just after the formation of hot dense neutron star the
conditions at the NS center (density $\sim 10^{15}~{\rm
g~cm^{-3}}$ and temperature $\sim 5\times 10^{11}~$K,
with a significant fraction of hyperons) allow for the nucleation a
quark matter nugget which then absorbs the whole neutron star
within minutes (Olinto 1987, Heiselberg \& Pethick
1993): a quark star is born. Conversion of baryonic matter into
SQM is strongly exothermic, releasing some $\sim 50~$MeV or
more per absorbed baryon. Most of that energy is lost in
neutrino -- anti-neutrino emission within fraction of a minute,
just as it happens with the initial gravitational collapse
when the hot neutron star had formed.
Conversion of the outer
layers of a neutron star into SQM is facilitated by strong
evaporation of nuclei into a neutron gas due to very high
temperature. The newborn quark star is a huge reservoir of
energy, including thermal energy of quarks,
which for simplicity are assumed to be normal
(non-superconducting) at the prevailing temperatures.
Quarks move freely within a
huge bag constituting the quark star but they cannot cross the
confining bag surface. However, electrons and
positrons, neutrinos, photons, and the magnetic field are not
subject to strong interaction; they can leave the quark bag
which plays the r{\^o}le of a membrane filtering
pure energy from hot quark matter. This outflow of energy
carrying zero baryon number can be a genuine progenitor of a
GRB associated with a SN Ic.

The duration of the pure energy outflow from hot,
rotating quark star is uncertain, and can be
seconds or many minutes, depending on the relative importance of
differential rotation and magnetic fields in the quark star
energy sharing.  The details of evolution of differentially
rotating quark star are beyond the scope of this paper.

{ The initial spectrum of GRB near the quark surface is very
specific (Usov 1998, 2001; Usov and Page 2002).} However, the
enormous optical depth modifies this original spectrum beyond
recognition.  We cannot provide any spectral information that
could be used as a direct diagnostic of quark star formation.

\section{Discussion and conclusions}
\label{sect:discuss}

Our suggestion that gamma-ray burst are related to the
formation of quark stars is mostly driven by the perceived
difficulty in generating bulk Lorenz factors as large as 300,
while at the same time keeping the energy density, and
therefore the optical depth very large.  Notice that critique
of the neutron-star -- quark-star conversion model of GRBs by
Fryer and Woosley (1998) does not apply to our scenario.
Namely, in our case the phase transition occurs on a timescale
much longer then the dynamic one, and therefore does not
provoke an outgoing shock, which would load energy outflow too
much with non-relativistic matter.

We cannot provide a proof that quark stars exist, but we have
a hint: the apparent bi-modality of SN Ic properties. Some
generate soft and weak bursts, like GRB980425/SN1998bw, with
no evidence for ultra-relativistic jets.  Some generate strong
gamma-ray bursts with ultra-relativistic jets, like
GRB030329/SN2003dh.  With pointing accuracy of {\it Swift},
supernovae like 1998bw will be readily detected, following the
breakout of a shock at the stellar surface.  The time to the
breakout is estimated to be approximately half a minute since
the core collapse (Deng 2005, private communication;  Woosley
2005, private communication). If the supernova generates a
relativistic jet in our direction, the soft precursor caused
by the breakout will be followed by a regular GRB a minute or
several minutes later, depending on poorly known details of a
transition from a neutron star to a quark star (Olinto 1987,
Heiselberg \& Pethick 1993). Swift should improve the
statistics of SN Ic, and the existence of precursors, and the
issue: is there a bimodal distribution of SNe Ic will be
observationally resolved.

Recent analysis of BATSE data by Lazzati (2005) indicates that
the precursors have already been detected, and if our interpretation
is correct they offer
evidence for the formation of quark stars following SN Ic
explosion.  The time interval between the onset of a precursor
and the beginning of the main GRB, $t_1 - t_2 $, can be used
as a diagnostics for a transition from a hot neutron star to a
hot quark star.

Needless to say, if the bi-modality is confirmed, we will be
strongly motivated to develop details of our model, including
the effects of quark superconductivity, and to use
quantitative description of stellar rotation, formation and
dissipation of magnetic field, as well as to describe the
energy transport  in a more detailed way.

While we consider the quark star solution to the problem of SN
Ic bi-modality, we realize this may be a result of rotation,
with high core angular velocity being responsible for the
relativistic jets.  The best argument we may offer in favor of
quark stars at this time are the time lags discovered by
Lazzati (2005), { provided the reality of precursors is
confirmed.}

The duration of GRB activity, observed to be up to several
minutes or more, may well be due to a gradual dissipation of a
differential rotation by the magnetic fields. The infall of
fresh nuclear matter following the initial core collapse may
provide not only an input of angular momentum but also a
supply of fresh baryons which will increase the mass of the
quark star.  Ultimately, the quark star may collapse into a
black hole, terminating the GRB activity.

{\it Swift} may provide information about the time lag between
the beginning of a strong GRB activity and an arrival of a
shock at the surface (the breakout, or a precursor).  With the
present state of a theory it is not even possible to predict
which may come first.  It will be certainly important to find
out observationally what is the relation between the
ultra-relativistic jet and the breakout of a more or less
spherical component of a SN Ic explosion.  Can the two
components be identified?  What is the time interval between
them? The distribution of long-lag GRBs found by Norris (2002)
indicates that these may be associated with supernovae like
1998bw, and {\it Swift} will be able to provide a definite
answer to the question: are these GRBs beamed away from us, or
are they approximately spherical explosions as envisioned by
Colgate (1968).

{It is somewhat disturbing that out of several dozen GRBs
detected with {\it Swift}, and half a dozen redshift
determinations (all with $ z > 1 $), not a single supernova
was detected. Given the dominance of long-lag GRBs at the
faint end of BATSE (Norris 2002), and their "Euclidean"
counts, some of those were expected to be relatively nearby,
and hence to produce easily detectable SNe Ic.}

While we cannot be quantitative in our description of GRB
energetics and spectra, it appears to us that quark-star
driven bursts are electromagnetic, rather than gas dynamical;
electromagnetic origin of GRBs was proposed by Blandford
(2002) and Lyutikov and Blandford (2003).

{ A problem with a standard GRB model, be it of a fireball or
an electromagnetic type, is the need to separate baryons from
energy.  This includes Kluzniak-Ruderman (1998) and Usov
(1992) models.  Note that  the strong magnetic fields are
generated with a dynamo process.  This can bring the field up
to equipartition with kinetic energy of moving baryons, and
makes the effective Lorentz factor only slightly larger than
1. It takes time to separate the field from the baryons.  It
is clear, that on a long enough time scale such a separation
happens, as demonstrated by magnetospheres of radio pulsars.
It is not obvious that the separation of magnetic energy from
baryons can be done on a time scale of seconds, and proceed up
to Lorentz factor $ \Gamma \sim 300 $. In our scenario we
begin with a baryon-free ultrarelativistic flow from a quark
star, and we have to avoid mixing it with contaminating
baryons.
Photons,  electron-positron pairs, and magnetic fields are
difficult to mix with baryons, and are likely to remain
separated.  In any case it is easier to maintain a
relativistic flow then to separate this flow from baryons.}

\section*{Acknowledgements}
We are very grateful to Dr. J. Deng and Dr. S. E. Woosley for
providing us with an estimate of the time interval to the
shock breakout at the surface of SN Ic.  We are most indebted
to Dr. D. Lazzati for bringing his analysis to our attention.
{ We are grateful to Dr. V.V. Usov for helpful remarks
concerning the photon spectrum of a newborn quark star.} We
apologize for not providing a complete list of references --
we tried to make sure that they are representative. This work
was partially supported by the Polish MNiI  grant no.
1P03D-008-27.

\bsp
\label{lastpage}

\end{document}